\documentclass[aps,pra,superscriptaddress,amsmath,amssymb,preprintnumbers]{revtex4}

\usepackage{amssymb}
\usepackage{graphicx}
\usepackage{graphics}
\usepackage{bm}
\usepackage{color}
\usepackage{cancel}
\usepackage{slashed}
\usepackage{subfigure}

\newcommand{\Ignore}[1]{}

\newcommand{\Ket}[1]{\left\vert #1\right\rangle}
\newcommand{\Bra}[1]{\left\langle #1\right\vert}

\newcommand{\ee}{\mathrm{e}}

\begin{document}

\title{Competition of Direct and Indirect Sources of Thermal Entanglement in a spin star network}

\author{Benedetto Militello}

\author{Anna Napoli}
\address{Dipartimento di Fisica e Chimica, Universit\`a degli Studi di Palermo, Via Archirafi 36, I-90123 Palermo, Italy}
\address{I.N.F.N. Sezione di Catania}

\begin{abstract}
A spin star system consisting of three peripheral two-state systems and a central one is considered, with the peripheral spins assumed to interact with each other, as well as with the central one. It is shown that such two couplings, each one being a thermal entanglement source, can significantly compete in the formation of quantum correlations in the thermal state, to the point that they can destroy any thermal entanglement of the peripheral spins.
\end{abstract}

\maketitle

\section{Introduction}\label{sec:introduction}

Temperature is usually considered an enemy of quantumness. Indeed,  when a system is in contact with a reservoir, it undergoes thermal fluctuations and loss of quantum coherence, which in turn usually produce a diminishing of quantum features. Nevertheless, even in the presence of strong thermal fluctuations or when the system has relaxed into its thermal state, quantum features can still be present. In caloritronics, temperature gradients are responsible for heat fluxes correlated, for example, to spin flowings (spintronics) in insulators~\cite{ref:Bauer2012,ref:Saf2017} or quasiparticle transport in flux qubits~\cite{ref:Zhao2003,ref:Zhao2004,ref:Spilla2015}. In Ref.~\cite{ref:Wu2011} it has been theoretically predicted a very counterintuitive enhancement of quantum correlations between two systems induced by the interaction with two different reservoirs at different temperatures. Temperature has also been predicted to be important for quantum Zeno effect. In fact, it has been theoretically proven that the border between quantum Zeno and anti-Zeno effect can be influenced by temperature~\cite{ref:Maniscalco2006}. Moreover, under certain conditions, temperature could be responsible for an enhancement of a generalized quantum Zeno effect induced by the coupling of a system with its environment~\cite{ref:MilitelloTZE}.  In this wide scenario, quantum features surviving in a thermal state, even at nonzero temperature, play an important role. 

Thermal entanglement, i.e., the degree of entanglement present in the thermal state of a compound system, has been introduced as a resource in quantum information processes and teleportation, since it allows to exploit quantum correlations still present in a system which has been thermalized~\cite{ref:Arnesen2001}. In recent studies, schemes of work extraction based on the \lq resource theory\rq\,have been proposed~\cite{ref:resourcetheory-1,ref:resourcetheory-2}, and, in such a context, entanglement between the system and an ancilla quantum system in a joint thermal state can be exploited to get work from the environment, which makes thermal entanglement useful also in the field of quantum thermodynamics~\cite{ref:quantumthermo}.

The existence of quantum correlations in thermalized systems has been explicitly connected with phase transitions~\cite{ref:Osterloh2002,ref:Osborne2002}. Quantum properties of physical systems connected with the production of thermal entanglement have been studied in spin chains described by Ising or Heisenberg models~\cite{ref:Wang2001,ref:Vedral2001,ref:Gong2009,ref:Zheng2017}, including the case of spins of different values~\cite{ref:Han2014}, in atom-cavity systems~\cite{ref:Wang2009}, in molecular models \cite{ref:Pal2010}. Applications to quantum teleportation have been proposed~\cite{ref:Zhang2007,ref:Zhou2009,ref:Fortes2017}. Moreover, nonclassical and nonlocal correlations in quantum systems relaxed toward their thermal states have been investigated~\cite{ref:Werlang2010,ref:Souza2009,ref:MingYong2012}.

Thermal entanglement has been studied in spin-star networks. For instance, Hutton and Bose~\cite{ref:Hutton2004} have analyzed the zero-temperature properties of such quantum systems, bringing to light interesting properties related to the parity of the number of outer (peripheral) spins. Wan-Li {\it et al} have studied the thermal entanglement in a spin-star network with three peripheral spins \cite{ref:Wan-Li2009}, evaluating pairwise entanglement between all possible couples of spins. In Refs~\cite{ref:Anza2010} and \cite{ref:Militello2010} a similar system has been analyzed, singling out tripartite thermal correlations by exploiting the tripartite negativity~\cite{ref:Sabin2008} and a genuine tripartite entanglement witness introduced by Huber {\it et al.}~\cite{ref:Huber2010}, respectively. By the way, it is worth mentioning that such two tools have produced extremely similar predictions concerning the tripartite thermal entanglement of the outer spins.

In this paper we consider a natural extension of the system analyzed in Refs.~\cite{ref:Anza2010,ref:Militello2010}, which consisted of three peripheral spins interacting with a central one (we call it \lq indirect\rq\, coupling). Here, we investigate a similar situation, but with the addition of another interaction between the peripheral spins (\lq direct\rq\, coupling). Each of such couplings is an entanglement source, and one could expect that they naturally cooperate to increase the amount of thermal entanglement of the outer spins. In spite of this expectation, we will see that these two thermal entanglement sources can compete, in some cases, driving the system to a thermal state that is less correlated than the thermal state obtained with only one of the two interactions. This effect will be demonstrated numerically and, in some parameter regions, analytically explained by looking at the eigenstates of the total Hamiltonian.

The paper is organized as follows. In the next section we introduce the model Hamiltonian and the mathematical tools for the analysis of the entanglement. In sec.~\ref{sec:tte} we analyze the behavior of the tripartite thermal entanglement in the parameter space, singling out the competition between the direct and indirect interactions. In sec.~\ref{sec:generalize} we extend our analysis to the case when the outer spins are more than three. Finally, in sec.~\ref{sec:conclusion} we give some conclusive remarks.

\section{The System}\label{sec:system}

{\it Model Hamiltonian --- } We start by considering a spin star system made of a central spin (labelled with \lq 0\rq) and $N=3$ peripheral spins (labelled with numbers from $1$ to $3$). We assume that each peripheral spin interacts with the central one and that, in addition, they interact with each other. We will also assume that all the natural frequencies are equal and that all the interaction terms conserve the total number of excitations. The relevant Hamiltonian can then be written as follows:
\begin{eqnarray}
\nonumber
H &=& \sum_{k=0}^{3} \frac{\omega_k}{2} \sigma_{k,z} + %
\sum_{k=1}^{3} ( \epsilon_k \sigma_{k,+}\sigma_{0,-} + \epsilon_k^* \sigma_{k,-}\sigma_{0,+} ) \\ %
&+& \sum_{k=1}^{3} ( \eta_k \sigma_{k,+}\sigma_{k+1,-} + \eta_k^* \sigma_{k,-}\sigma_{k+1,+} ) %
\end{eqnarray}
where $\sigma_{k,\alpha}$, with $\alpha=z,\pm$, are the Pauli operators, $\epsilon_k$ and $\eta_k$ are coupling constants and where we have assumed the notation $\sigma_{4,\alpha}=\sigma_{1,\alpha}$.

For the sake of simplicity we will assume invariance of the system under rotation of an angle $2\pi/3$, which means that the three peripheral spins are considered as perfectly equal, hence implying $\epsilon_k=\epsilon_j$ and $\eta_k=\eta_j$, $\forall k,j$. Moreover, we will assume real coupling constants ($\epsilon_k=\epsilon_k^*$ and $\eta_k=\eta_k^*$, $\forall k$), so that the model simplifies as:
\begin{eqnarray}
\nonumber
H &=& \sum_{k=0}^{3} \frac{\omega}{2} \sigma_{k,z} + %
\sum_{k=1}^{3} \epsilon ( \sigma_{k,+}\sigma_{0,-} + \sigma_{k,-}\sigma_{0,+} ) \\ %
&+& \sum_{k=1}^{3} \eta ( \sigma_{k,+}\sigma_{k+1,-} + \sigma_{k,-}\sigma_{k+1,+} )\,. %
\end{eqnarray}

This Hamiltonian conserves the total number of excitations: $N=4\mathbf{I}+\sum_k \sigma_{k,z}=\sum_k \Ket{1}_k\Bra{1}$, with $\sigma_{k,z}\Ket{1}_k=\Ket{1}_k$ (and $\sigma_{k,z}\Ket{0}_k=-\Ket{0}_k$).

{\it Witnessing Tripartite Entanglement --- } Since we want to analyze tripartite entanglement (in reduced thermal states obtained after tracing over the central spin degrees of freedom), we have to choose which tool is preferable to detect it.

Generally speaking, detecting and quantifying entanglement is a though job. It is well known how to identify it for bipartite systems~\cite{ref:Peres,ref:Zyczkowski,ref:Vidal}, and, if the system has only few degrees of freedom, it is also known how to quantify it, even for mixed states~\cite{ref:Wootters}. Tripartite and multipartite systems are even more difficult to analyze~\cite{ref:Fazio2008}. A successful tool that allows to quantify tripartite entanglement in pure states is the three-tangle~\cite{ref:ThreeTangle}, which, in spite of raised criticisms~\cite{ref:DoesThreeTangle}, is widely used. As previously mentioned, in 2010 Huber {\it et al.} have introduced an entanglement witness able to reveal the presence of genuine tripartite entanglement, even in mixed states. Another effective witness of tripartite entanglement, widely used and valid for mixed states, is the tripartite negativity, that we will use to develop our analysis.

The negativity \cite{ref:Peres,ref:Zyczkowski,ref:Vidal} of the state of a quantum system made of the two parts $\mathrm{A}$ and $\mathrm{B}$ is expressed as:
\begin{equation}
\mathcal{N}_{\mathrm{A}-\mathrm{B}} (\tau) = \sum_{i}
|\lambda_i(\tau^{T\mathrm{A}})| - 1 \quad ,
\end{equation}
where $\lambda_i(\tau^{T\mathrm{A}})$ is the $i$-th eigenvalue of $\tau^{T\mathrm{A}}$, which is the partial transpose related to subsystem $\mathrm{A}$ of the total ($\mathrm{A}-\mathrm{B}$) density matrix $\tau$.

For a tripartite system $\mathrm{A}-\mathrm{B}-\mathrm{C}$ it is possible to introduce three different negativities, related to the three possible bipartitions: $\mathcal{N}_{\mathrm{A}-\mathrm{B}\mathrm{C}}$ (which considers $\mathrm{B}$ and $\mathrm{C}$ as a whole), $\mathcal{N}_{\mathrm{B}-\mathrm{A}\mathrm{C}}$ (which considers $\mathrm{A}$ and $\mathrm{C}$ as a whole), and $\mathcal{N}_{\mathrm{C}-\mathrm{A}\mathrm{B}}$ (which considers $\mathrm{A}$ and $\mathrm{B}$ as a whole). The geometric average of such three bipartite negativities is said the tripartite negativity of the system. Therefore, in our case, having spins $1$, $2$ and $3$, we can introduce the following:
\begin{equation} \label{eq:neggeom}
\mathcal{N}(\tau) =
\sqrt[3]{\mathcal{N}_{1-23}(\tau)\,\mathcal{N}_{2-13}(\tau)\,\mathcal{N}_{3-12}(\tau)}\;.
\end{equation}
This quantity has some interesting properties which have been proven in Ref.~\cite{ref:Sabin2008}:
\begin{itemize}
\item[i)] if $\tau$ is separable or simply bi-separable then $\mathcal{N} = 0$; %
\item[ii)] $\mathcal{N}$ is invariant under Local Unitary operators; %
\item[iii)] $\mathcal{N}$ is non increasing under Local Operations and Classical Communication (LOCC) operators.
\end{itemize}

It is worth remarking that $\mathcal{N}\not=0$ is not a sufficient condition to single out the presence of tripartite entanglement. Nevertheless, a non vanishing tripartite negativity implies that none of the three subsystems is separable, hence revealing quantum correlations involving all the three subsystems.

{\it Thermal Entanglement --- } Since we are investigating the tripartite thermal entanglement of the peripheral two-state systems, we evaluate the thermal state of the whole system, we trace over the central spin degrees of freedom and then evaluate the degree of tripartite entanglement between the three peripheral spins. The reduced thermal state of the outer spins is simply given by:
\begin{equation} \label{eq:thermalstate} 
\rho_{123} = \mathrm{tr}_{0} \ee^{- H / (k_\mathrm{B}T)}\,,
\end{equation}
with $k_\mathrm{B}$ the Boltzmann constant and $T$ the temperature the system has thermalized to.

\section{Analysis of Thermal Entanglement}\label{sec:tte}

\begin{figure}
\subfigure[]{\includegraphics[width=0.40\textwidth, angle=0]{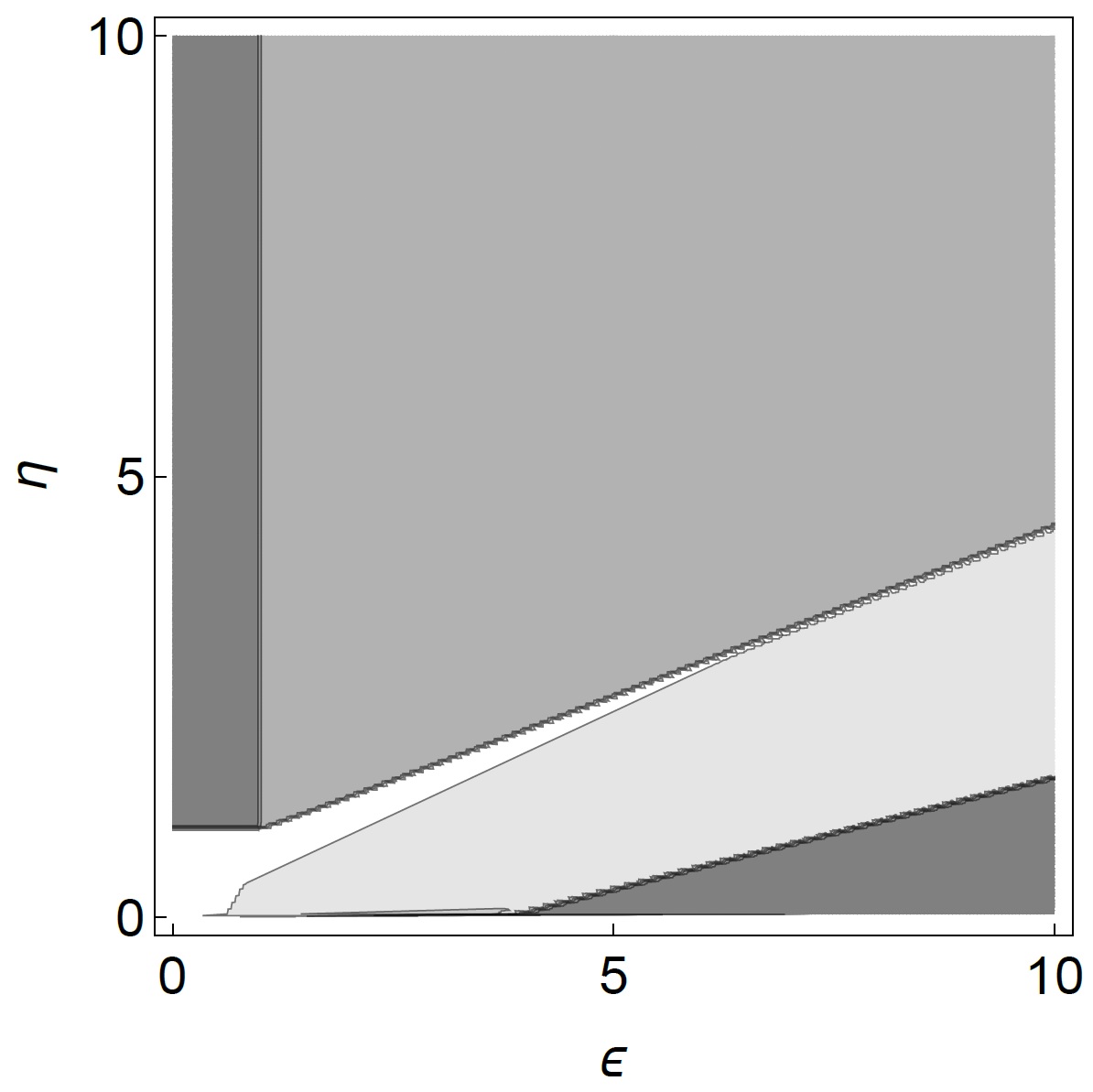}} \qquad
\subfigure[]{\includegraphics[width=0.40\textwidth, angle=0]{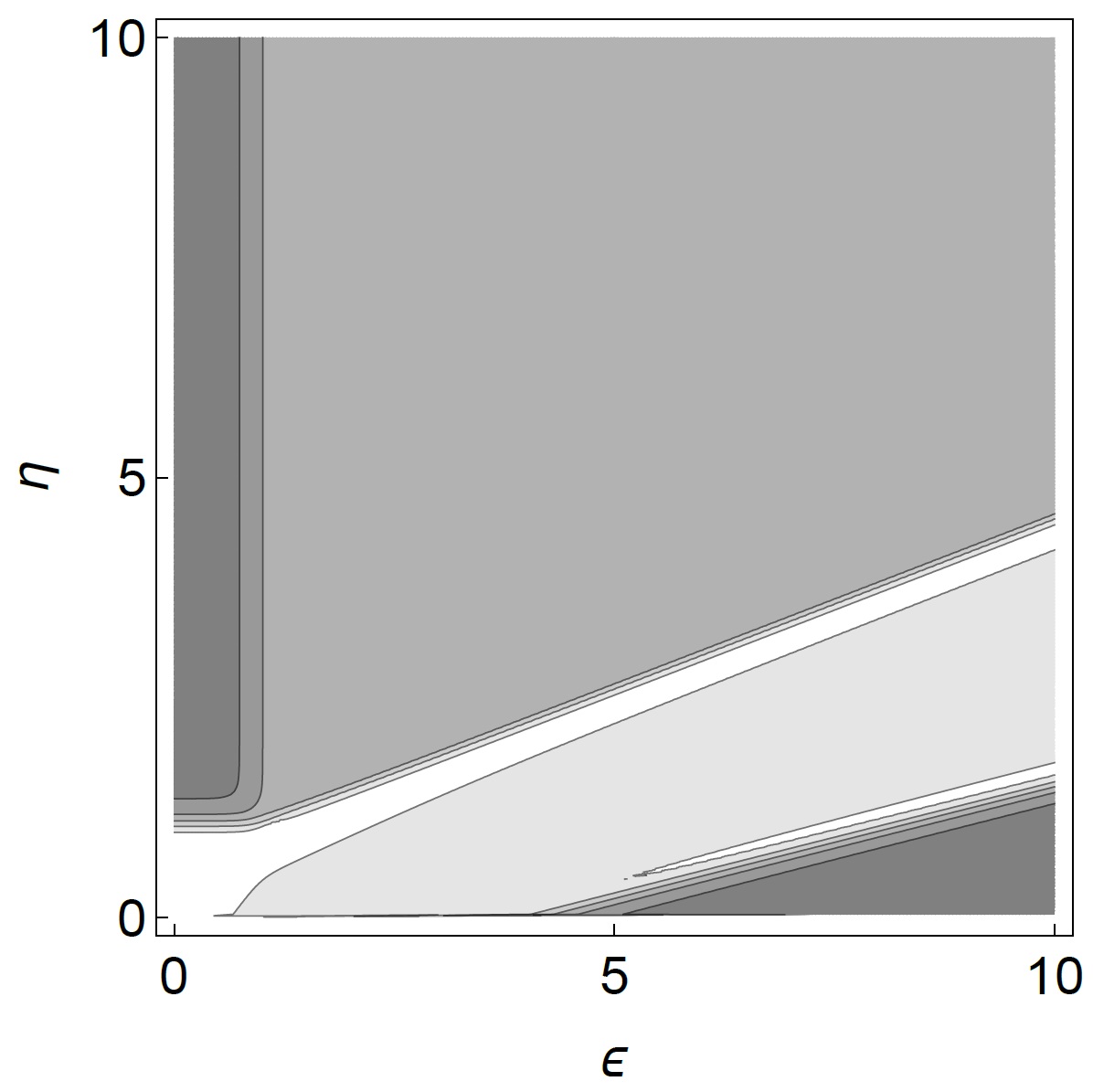}}\\
\subfigure[]{\includegraphics[width=0.40\textwidth, angle=0]{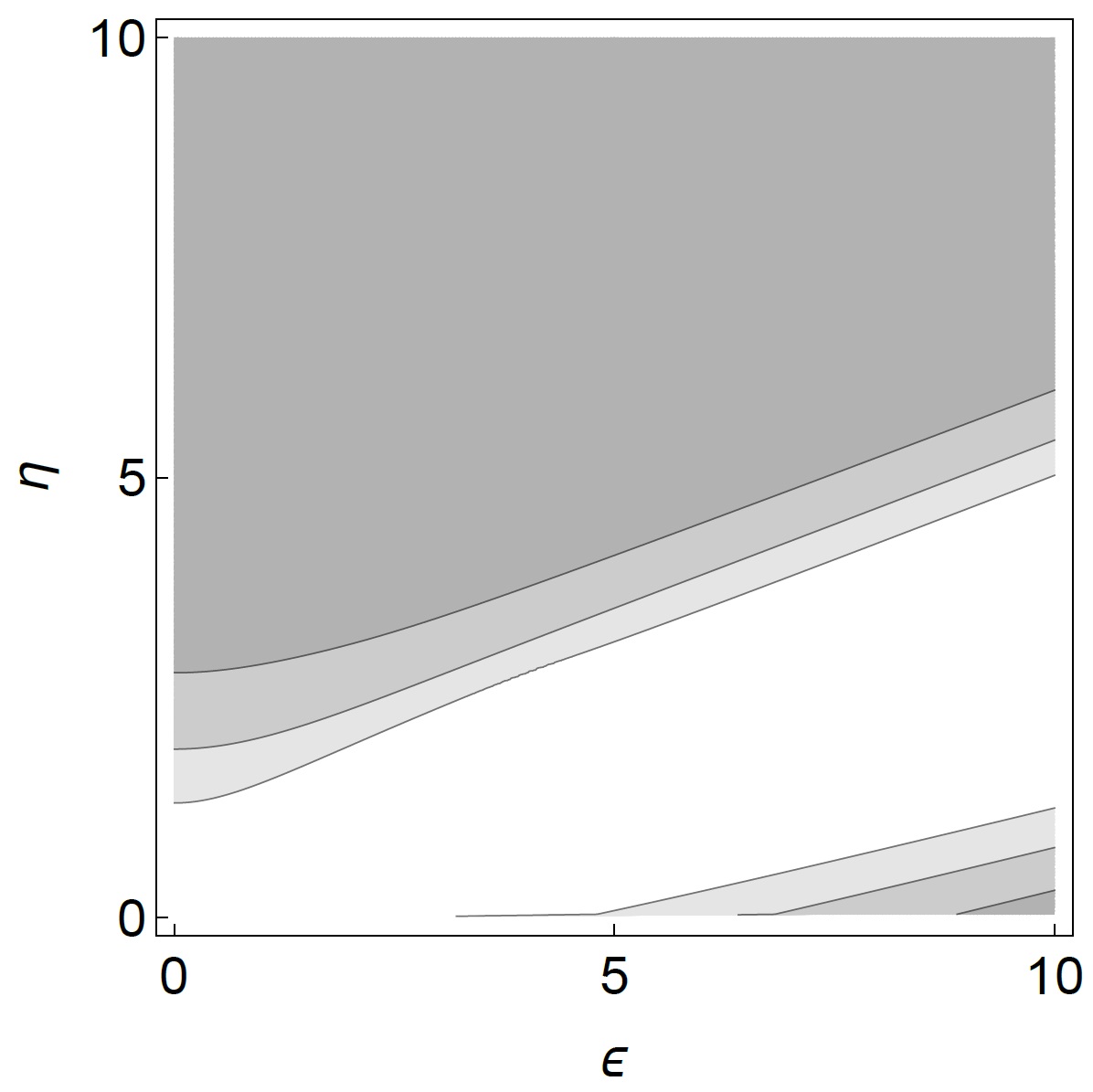}} \qquad
\subfigure[]{\includegraphics[width=0.40\textwidth, angle=0]{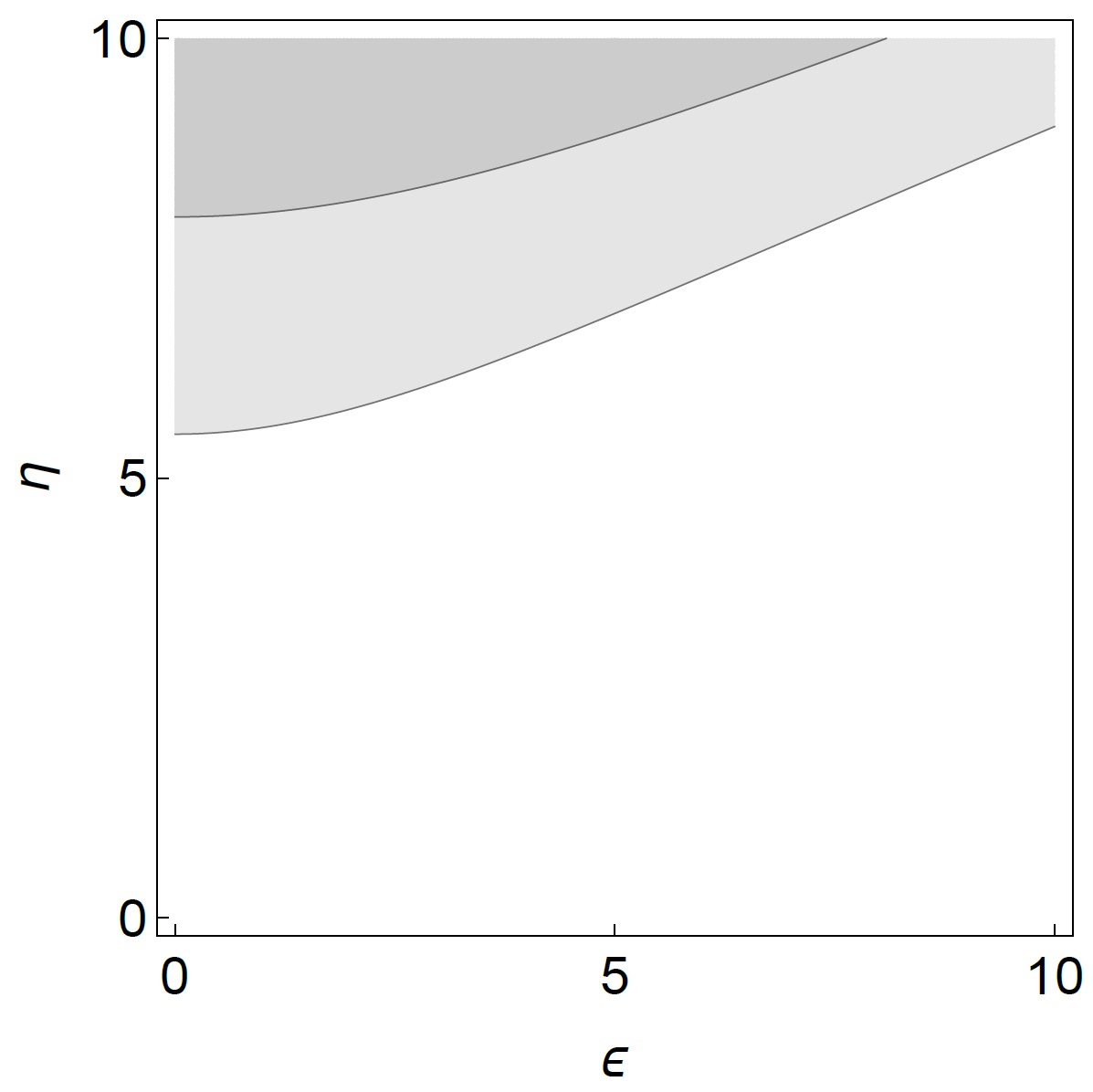}}
\includegraphics[width=0.08\textwidth, angle=0]{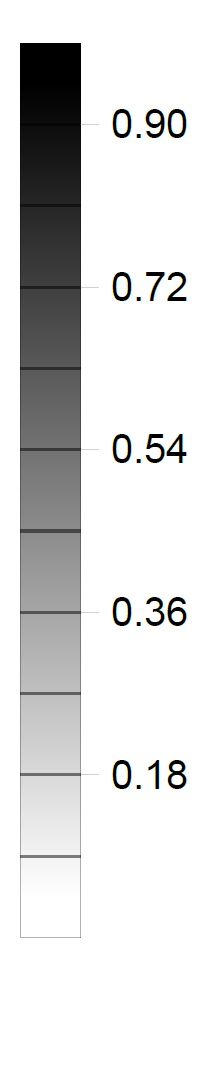}
\caption{Tripartite negativity $\mathcal{N}$ of the reduced thermal state of the outer spins as a function of $\epsilon$ and $\eta$ (both in units of $\omega$), for $k_\mathrm{B} T / (\hbar\omega) = 0.01$ (a), $k_\mathrm{B} T / (\hbar\omega) = 0.1$ (b), $k_\mathrm{B} T / (\hbar\omega) = 1$ (c), $k_\mathrm{B} T / (\hbar\omega) = 5$ (d). (A darker grey corresponds to a higher value.)} \label{fig:TotalBehaviour}
\end{figure}

\begin{figure}
\subfigure[]{\includegraphics[width=0.45\textwidth, angle=0]{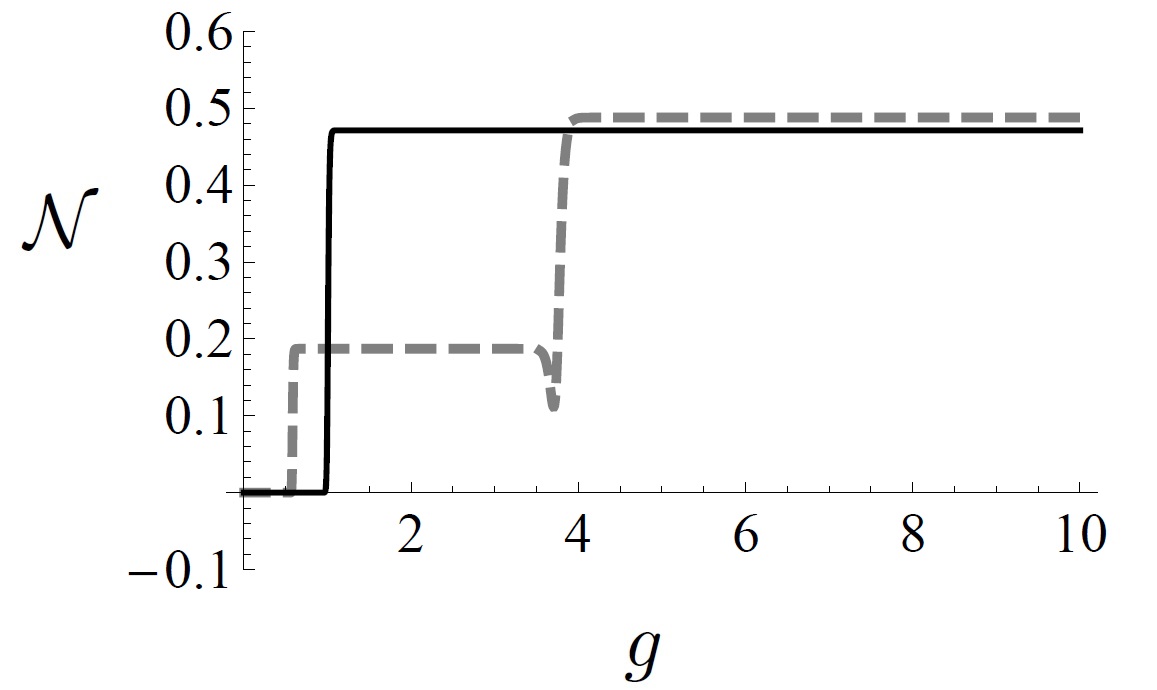}} \qquad%
\subfigure[]{\includegraphics[width=0.45\textwidth, angle=0]{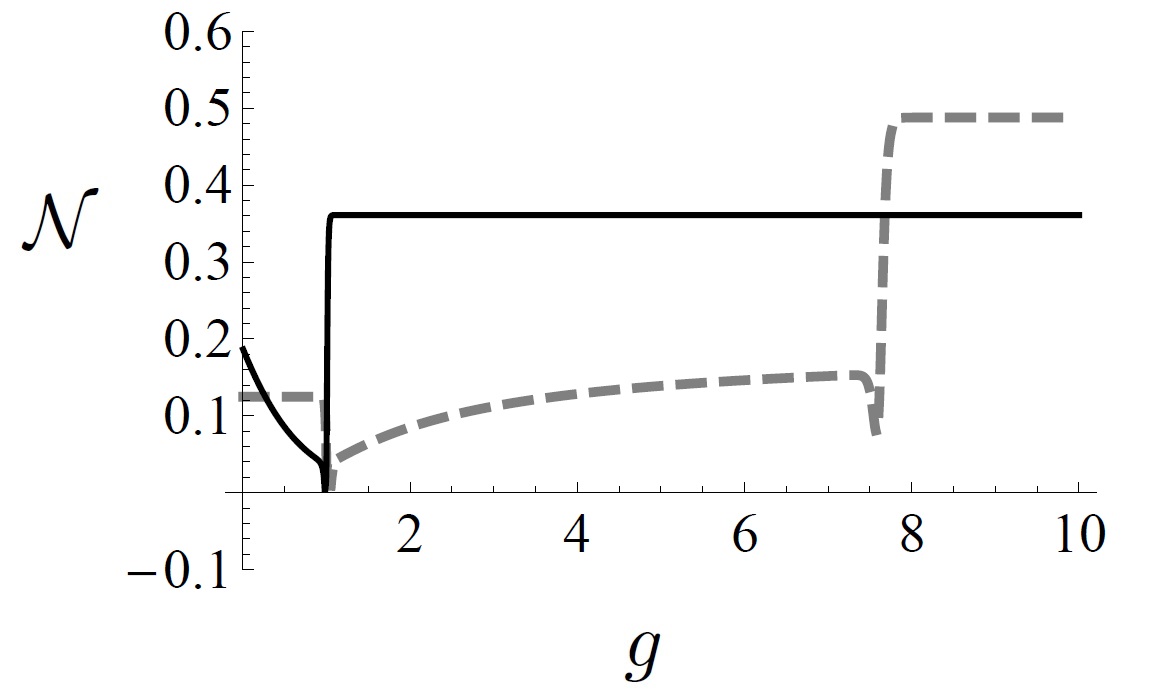}} %
\caption{Tripartite negativity $\mathcal{N}$ of the reduced thermal state of the outer spins at small temperature ($k_\mathrm{B} T / (\hbar\omega) = 0.01$) for different values of the coupling constants ($g$ indicates the generic coupling constant): in (a) $\mathcal{N}$ as a function of $\eta$ for $\epsilon=0$ (black solid line) and $\mathcal{N}$ as a function of $\epsilon$ for $\eta=0$ (grey dashed line); in (b) $\mathcal{N}$ as a function of $\eta$ for $\epsilon/\omega=1$ (black solid line) and $\mathcal{N}$ as a function of $\epsilon$ for $\eta/\omega=1$ (grey dashed line). (Both $\eta$ and $\epsilon$ are given in units of $\omega$.)} \label{fig:ZeroTemp}
\end{figure}

\begin{figure}
\subfigure[]{\includegraphics[width=0.45\textwidth, angle=0]{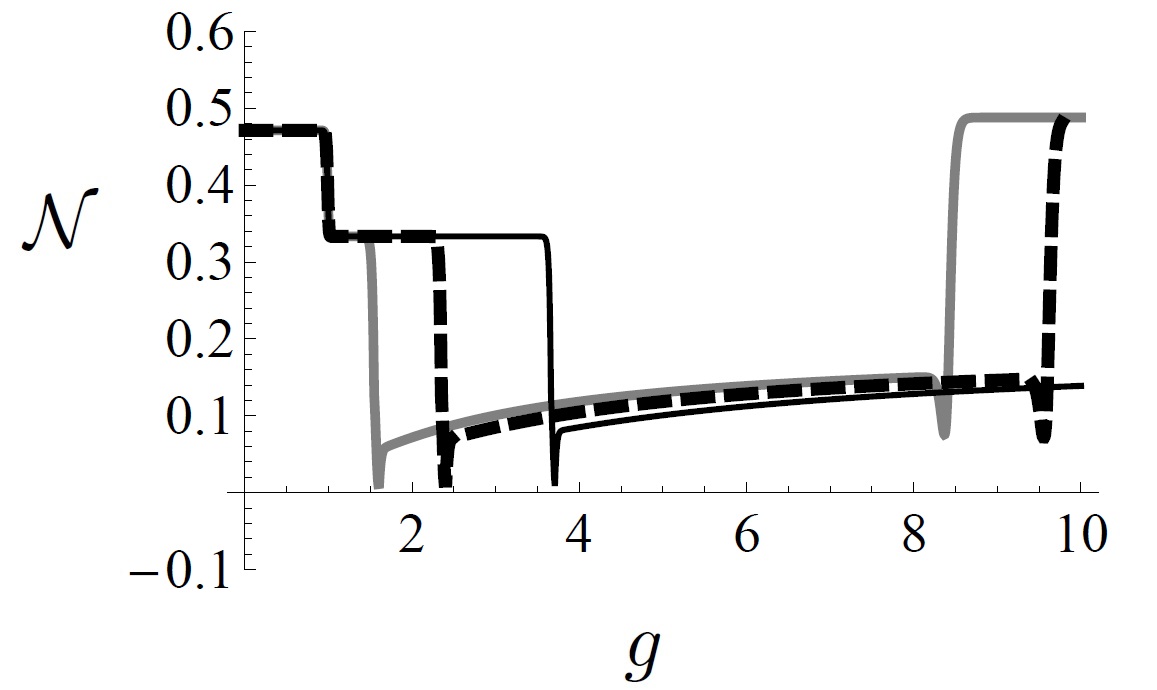}} \qquad%
\subfigure[]{\includegraphics[width=0.45\textwidth, angle=0]{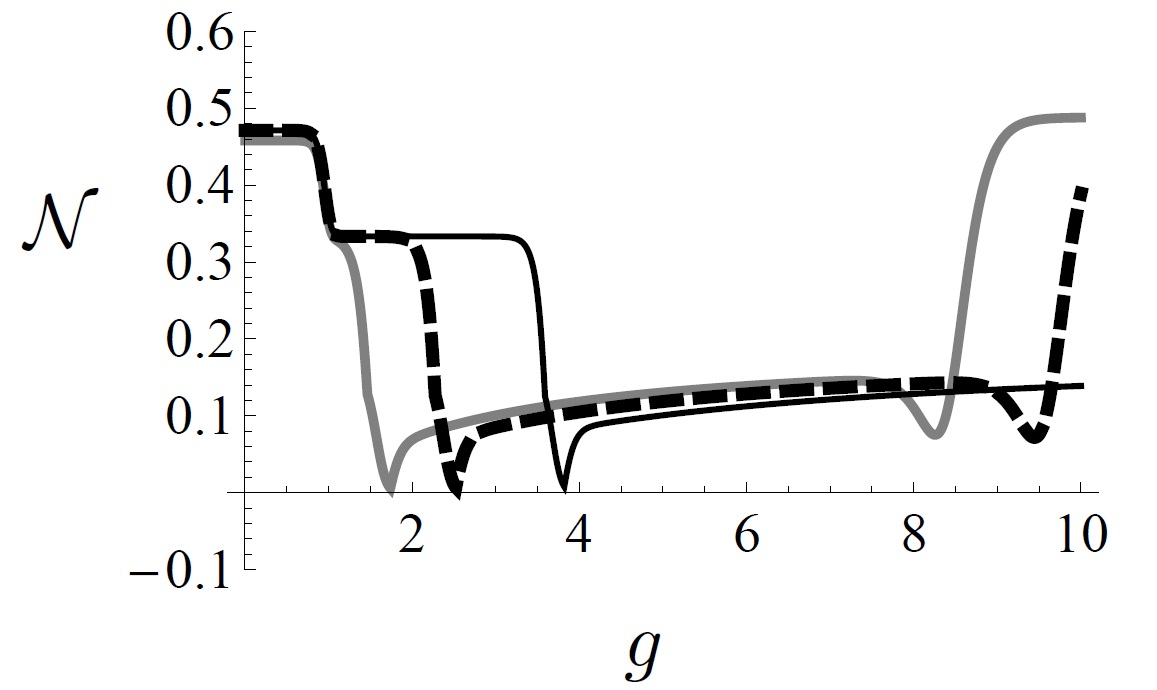}} \\%
\subfigure[]{\includegraphics[width=0.45\textwidth, angle=0]{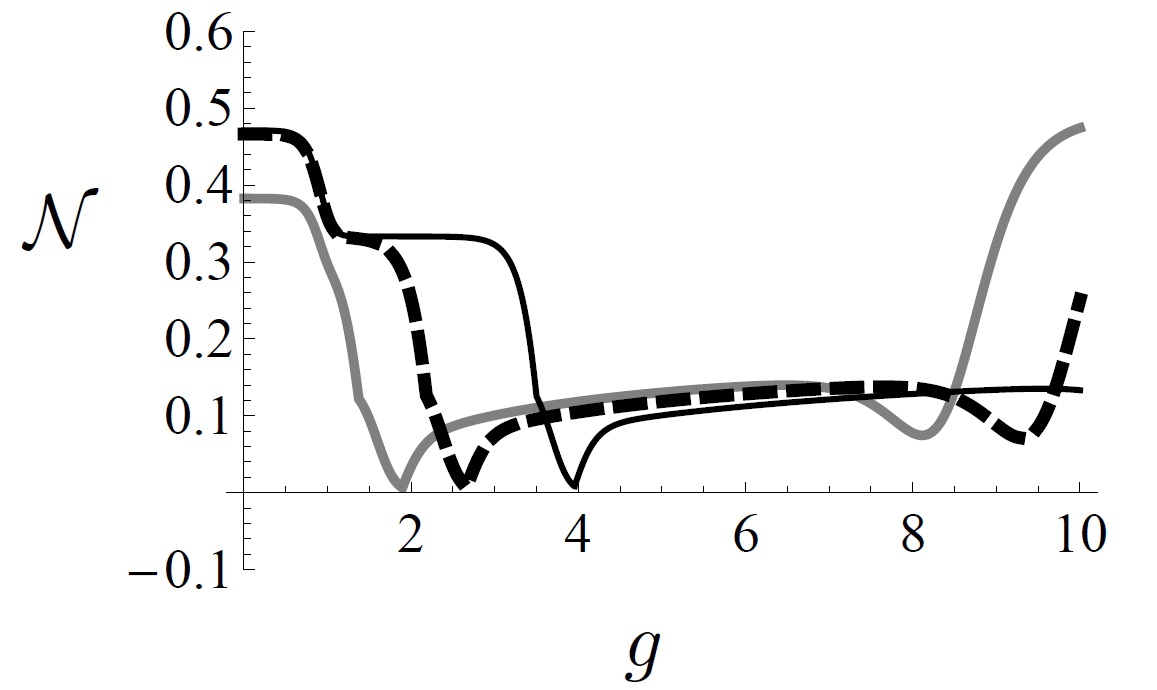}} \qquad%
\subfigure[]{\includegraphics[width=0.45\textwidth, angle=0]{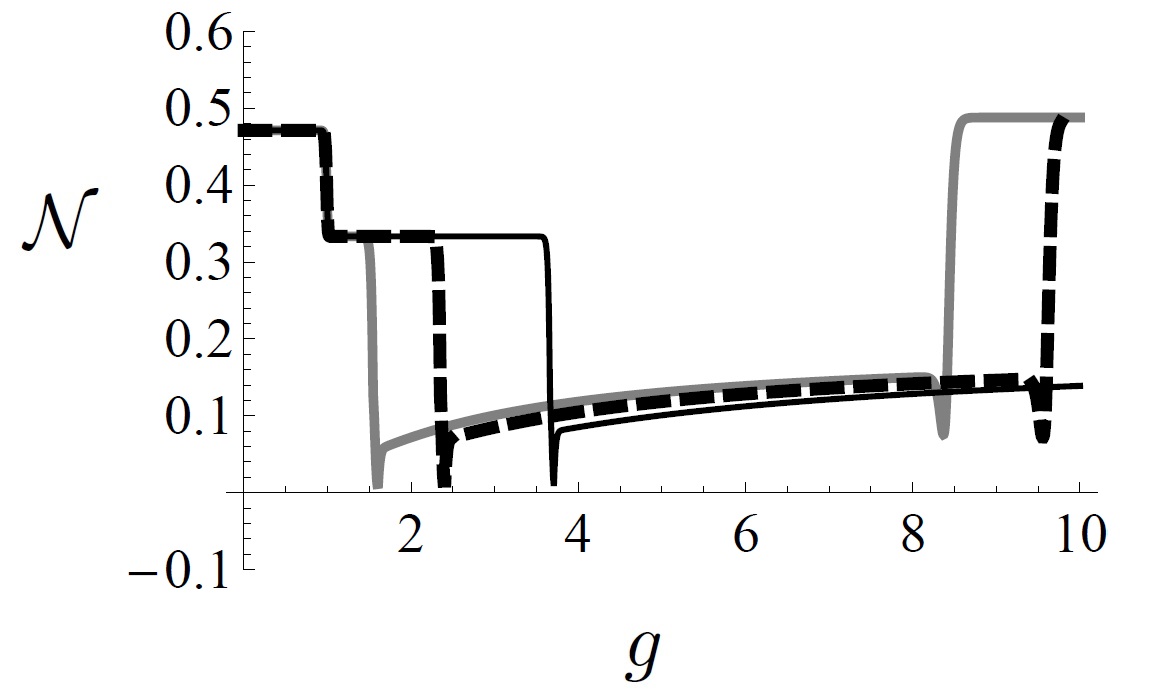}} %
\caption{Tripartite negativity $\mathcal{N}$ of the reduced thermal state of the of the outer spins at different temperature for different values of the coupling constants: $\mathcal{N}$ as a function of $\epsilon$ (in units of $\omega$) for $\eta/\omega=1.2$ (grey bold solid line), $\eta/\omega=1.5$ (black bold dashed line), $\eta/\omega=2$ (black thin solid line). The values of temperature considered are: $k_\mathrm{B} T / (\hbar\omega) = 0.01$ (a), $k_\mathrm{B} T / (\hbar\omega) = 0.05$ (b), $k_\mathrm{B} T / (\hbar\omega) = 0.1$ (c), $k_\mathrm{B} T / (\hbar\omega) = 0.5$ (d).}
\label{fig:IncreasingTemperature}
\end{figure}

\begin{figure}
\subfigure[]{\includegraphics[width=0.45\textwidth, angle=0]{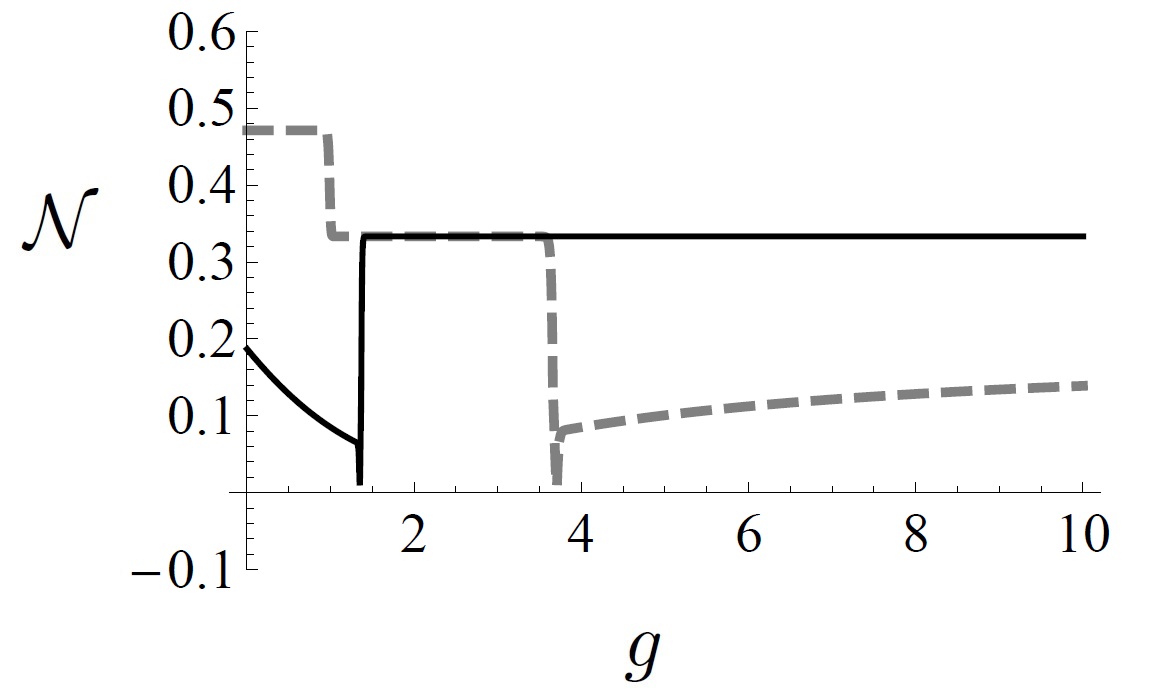}} \qquad%
\subfigure[]{\includegraphics[width=0.45\textwidth, angle=0]{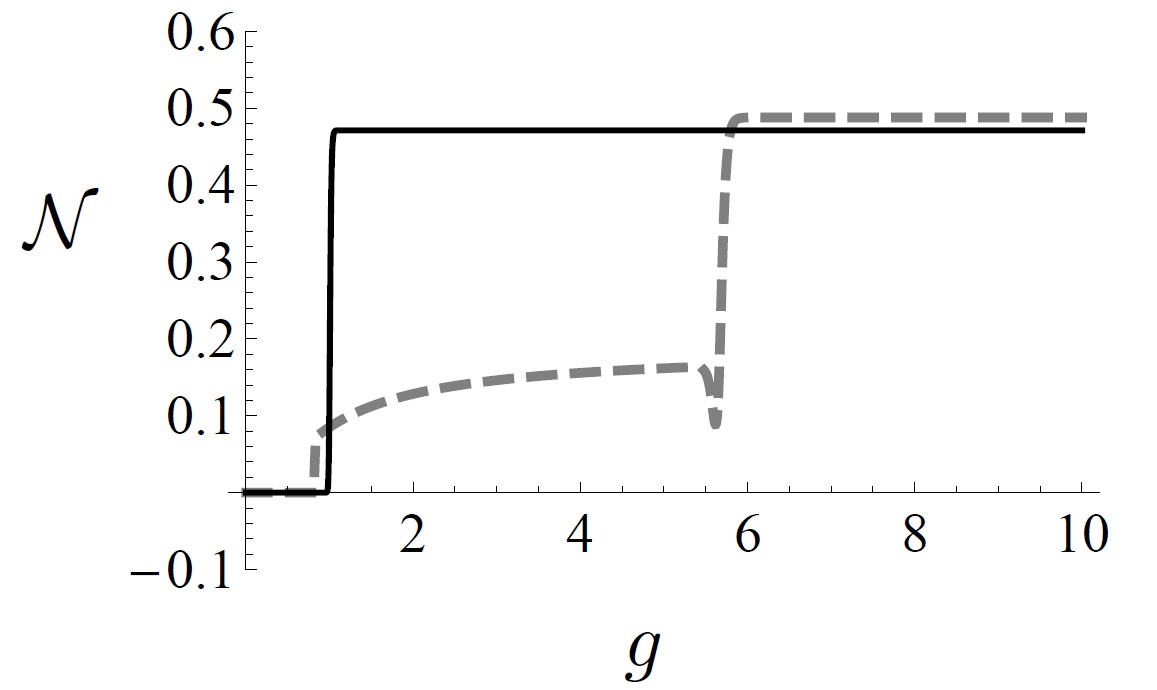}} %
\caption{Tripartite thermal negativity $\mathcal{N}$ of the reduced thermal state of the of the outer spins at small temperature ($k_\mathrm{B} T / (\hbar\omega) = 0.01$) for different values of the coupling constants ($g$ indicates the generic coupling constant, $\epsilon$ or $\eta$): in
(a) $\mathcal{N}$ as a function of $\eta$ (in units of $\omega$) for $\epsilon/\omega=2$ (black solid line) and $\mathcal{N}$ as a function of $\epsilon$ (in units of $\omega$) for $\eta/\omega=2$ (grey dashed line); in (b) $\mathcal{N}$ as a function of $\eta$ for $\epsilon/\omega=0.5$ (black solid line) and $\mathcal{N}$ as a function of $\epsilon$ for $\eta/\omega=0.5$ (grey dashed line).}
\label{fig:OmegaNoEpsilon}
\end{figure}

Even though persistence of thermal entanglement at high temperature has been predicted under special conditions~\cite{ref:Souza2009,ref:Ferrera2006}, thermal entanglement typically diminishes when temperature increases, at least in the limit of very high temperature. This is due to the fact that at high temperature all the states are almost equally populated, which makes the density operator of the system more and more separable. At low and relatively low temperature, residual entanglement could still be present, due to the fact that the states with lower energies are more populated than the others. Of course, we find these pretty general traits in our system, as well. Nevertheless, we have found some specific behavior related to the competition between couplings. Figs.~\ref{fig:TotalBehaviour}, \ref{fig:ZeroTemp}, \ref{fig:IncreasingTemperature} and \ref{fig:OmegaNoEpsilon}  show  the behavior of the thermal entanglement in our system for different values of the coupling constants $\eta$ and $\epsilon$.

In Fig.~\ref{fig:TotalBehaviour} it is shown the amount of tripartite thermal entanglement for the three peripheral spins as a function of the two coupling constants $\epsilon$ and $\eta$, for different temperatures.  It is well visible that for lower temperatures the amount of entanglement is generally higher, as expected. In fact, for higher temperatures the white area (zero or almost zero entanglement) becomes wider.

One could expect that a stronger coupling, whether direct (between the peripheral spins, then related to $\eta$) or indirect (between the central spin and the peripheral ones, then related to $\epsilon$), would produce a higher amount of entanglement. Moreover, one could expect that if both the coupling constants increase then more and more entanglement will be left to the system when it is in the thermal state. In other words, one expects that two sources of tripartite entanglement will cooperate
in the production of tripartite thermal entanglement of the peripheral spins. However, an accurate observation of the first of the four plots of Fig.~\ref{fig:TotalBehaviour}  makes one argue that fixing a coupling constant, the amount of entanglement could
be non monotonic with respect to the other coupling constant. See for example that in Fig.~\ref{fig:TotalBehaviour}a, corresponding to virtually zero temperature ($k_\mathrm{B}T/\hbar\omega=0.01$), the line individualized by $\eta/\omega\approx 1$ and
$\epsilon/\omega$ varying from $0$ to $10$ intercepts a dark gray zone, a lighter gray zone, the white zone, another light gray zone and another dark gray zone, meaning that for increasing $\epsilon$ one has a diminishing of entanglement first and then an increase. Something similar happens for $\epsilon/\omega>5$ and $\eta/\omega$ spanning from $0$ to $10$. Though less evident and significant, a degree of non monotonicity is still present at a higher temperature, as one can see in Figs.~\ref{fig:TotalBehaviour}b ($k_\mathrm{B}T/\hbar\omega=0.1$) and \ref{fig:TotalBehaviour}c ($k_\mathrm{B}T/\hbar\omega=1$). For $k_\mathrm{B}T/\hbar\omega=5$ there is no evidence of non monotonicity for the ranges of values of $\epsilon$ and $\eta$ considered in Fig.~\ref{fig:TotalBehaviour}d.

Since the non monotonicity is more visible at zero temperature, in Fig.~\ref{fig:ZeroTemp} the behaviors of peripheral thermal entanglement at virtually zero temperature is considered. In Fig.~\ref{fig:ZeroTemp}a the zero temperature entanglement is considered when only one of the two couplings is present. Instead, in Fig.~\ref{fig:ZeroTemp}b the behavior with both the couplings present is considered. In particular, one of the two coupling strengths is fixed while the other spans an extended range of values. A non monotonic behavior is very well visible in the black solid line, corresponding to $\epsilon/\omega=1$ and $\eta/\omega$ spanning from $0$ to $10$.

Fig.~\ref{fig:IncreasingTemperature} show other non monotonic behaviors for different values of the direct interaction strength $\eta$ (different values in each plot) and for different temperatures: $k_\mathrm{B} T / (\hbar\omega) = 0.01$ (\ref{fig:IncreasingTemperature}a), $k_\mathrm{B} T / (\hbar\omega) = 0.05$ (\ref{fig:IncreasingTemperature}b), $k_\mathrm{B} T / (\hbar\omega) = 0.1$ (\ref{fig:IncreasingTemperature}c), $k_\mathrm{B} T / (\hbar\omega) = 0.5$ (\ref{fig:IncreasingTemperature}d). At virtually zero temperature (\ref{fig:IncreasingTemperature}a) sharp changes of the amount of entanglement which resemble phase transitions are well visible, but when the temperature is not virtually zero (\ref{fig:IncreasingTemperature}b, \ref{fig:IncreasingTemperature}c and \ref{fig:IncreasingTemperature}d) the curves become smoother, though the non-monotonicity is still well visible.

The general behavior of the thermal entanglement is difficult to explain, since it involves the complete structure of the eigenstates of the Hamiltonian, the weights of them in the thermal state given by the Boltzmann factors and, in the considered case,
the effect of tracing over the central spin variables. Understanding the thermal entanglement at zero temperature is instead easier since it coincides with understanding the properties of the ground state of the system. Now, the Hamiltonian
we have considered has the following eigenvalues: $\eta - \sqrt{3 \epsilon^2 + \eta^2} - \omega$, $-2 \omega$, $-\epsilon - \eta$, $-\epsilon - \eta$, $-2 (\epsilon - \eta)$, $\epsilon - \eta$, $\epsilon - \eta$, $2 (\epsilon + \eta)$, $-\eta - \omega$, $-\eta
- \omega$, $\eta + \sqrt{3 \epsilon^2 + \eta^2} - \omega$, $2 \omega, -\eta + \omega$, $-\eta + \omega$, $\eta - \sqrt{3 \epsilon^2 + \eta^2} + \omega$, $\eta + \sqrt{3 \epsilon^2 + \eta^2} + \omega$. It is easy to see that for $\epsilon=\omega$ and $0<\eta<\omega$ the two lowest eigenvalues are the first and second of the list, i.e., $\eta - \sqrt{3 \epsilon^2 + \eta^2} - \omega$ (the lowest) and $-2\omega$ (the second lowest), which correspond to the eigenstates $\Ket{\psi_1}\propto(\eta+ \sqrt{3\epsilon^2+\eta^2})\Ket{1}_0 \Ket{000}_{123} - \epsilon \Ket{0}_0 (\Ket{100}_{123}+\Ket{010}_{123}+\Ket{001}_{123})$  and $\Ket{\psi_{2}}=\Ket{0}_0 \Ket{000}_{123}$, respectively (with the obvious notation that $\Ket{\phi}_0$ indicates a state of the central spin and $\Ket{\varphi}_{123}$ a state of the three peripheral ones). The state $\Ket{\psi_1}$, which is the ground state for $\eta<\omega$, contains a Werner state of the three peripheral spin, giving a certain amount of entanglement. Nevertheless, as $\eta$ becomes higher, the relative weight of the coefficient of $\Ket{1}_0 \Ket{000}_{123}$ becomes higher and the amount of thermal entanglement diminish because of the direct interaction between the peripheral spins. When $\eta$ reaches the values $\omega$ the two lowest levels cross. At the same time also the two energies $-\epsilon-\eta$ (a doublet) and $-\eta-\omega$ (another doublet) reach the energy of the ground state, so that six levels cross at the same time. For $\eta>\omega$, the two doublets $-\epsilon-\eta$ and $-\eta-\omega$ constitute a four-degenerate ground state and, since the corresponding eigenstates are independent from $\eta$ and $\epsilon$, a plateaux is present in the relevant entanglement plot.

In Fig.~\ref{fig:OmegaNoEpsilon} we show more non monotonicity at zero temperature for $\epsilon/\omega=2$ and varying $\eta$ (as well as $\eta/\omega=2$ and varying $\epsilon$), and for $\epsilon/\omega=0.5$ and varying $\eta$ (as well as $\eta/\omega=0.5$ and varying $\epsilon$). Although details can be different from the previous plots, diminishing and subsequent increasing of thermal entanglement for increasing coupling constants are still present.

\section{Increasing the size of the outer circle}\label{sec:generalize}

It could be of interest to consider an extension of the previous model to the case of $M>3$ outer spins. Concerning the interactions, one can consider couplings between each pair of the outer spins, or, as a simplification, interactions between next neighbor outer spins only. In the latter case, the relevant Hamiltonian is given by:  
\begin{eqnarray}
\nonumber
H &=& \sum_{k=0}^{M} \frac{\omega}{2} \sigma_{k,z} + %
\sum_{k=1}^{M} \epsilon (\sigma_{k,+}\sigma_{0,-} + \sigma_{k,-}\sigma_{0,+} ) \\ %
&+& \sum_{k=1}^{M} \eta (\sigma_{k,+}\sigma_{k+1,-} + \sigma_{k,-}\sigma_{k+1,+} )\,, %
\end{eqnarray}
where we assume the notation $\sigma_{M+1,\pm} = \sigma_{1,\pm}$.

In order to study the thermal entanglement of the outer spins, we introduce a natural generalization of the tripartite negativity (for which we still use the symbol $\mathcal{N}$) as follows:
\begin{eqnarray}
\mathcal{N}(\tau) = \sqrt[M]{\prod_{n=1}^M \mathcal{N}_{n - \overline{n}}(\tau)} \,,
\end{eqnarray}
where $\overline{n}$ is the set of integer values $\{1,2,..., \overline{n}-1, \overline{n}+1,..., M\}$ and $\mathcal{N}_{n - \overline{n}}$ indicates the bipartite negativity between the $n$-th spin and the rest of the outer spins considered as a whole.

Though there is not a theoretical analysis of the properties of this parameter for $M>3$, it is clear that the main property still holds: when the parameter $\mathcal{N}$ is non zero none of the outer spins is separable from the other $M-1$. By the way, the other properties, i.e., invariance under local unitary operations and its being non-increasing under LOCC operators, are easily proven in the same way they can be proven for the tripartite negativity.

\begin{figure}
\subfigure[]{\includegraphics[width=0.40\textwidth, angle=0]{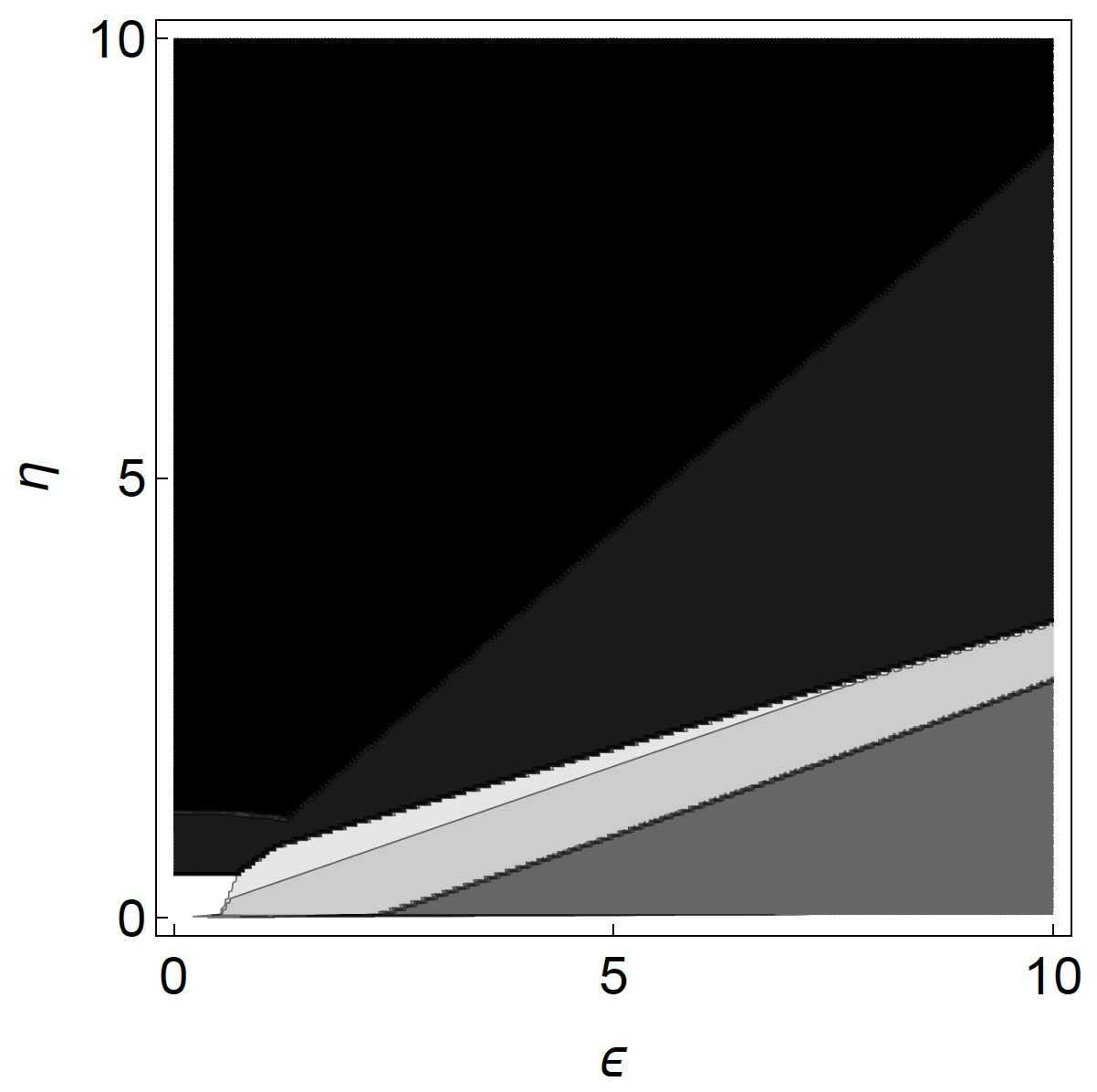}} \qquad
\subfigure[]{\includegraphics[width=0.40\textwidth, angle=0]{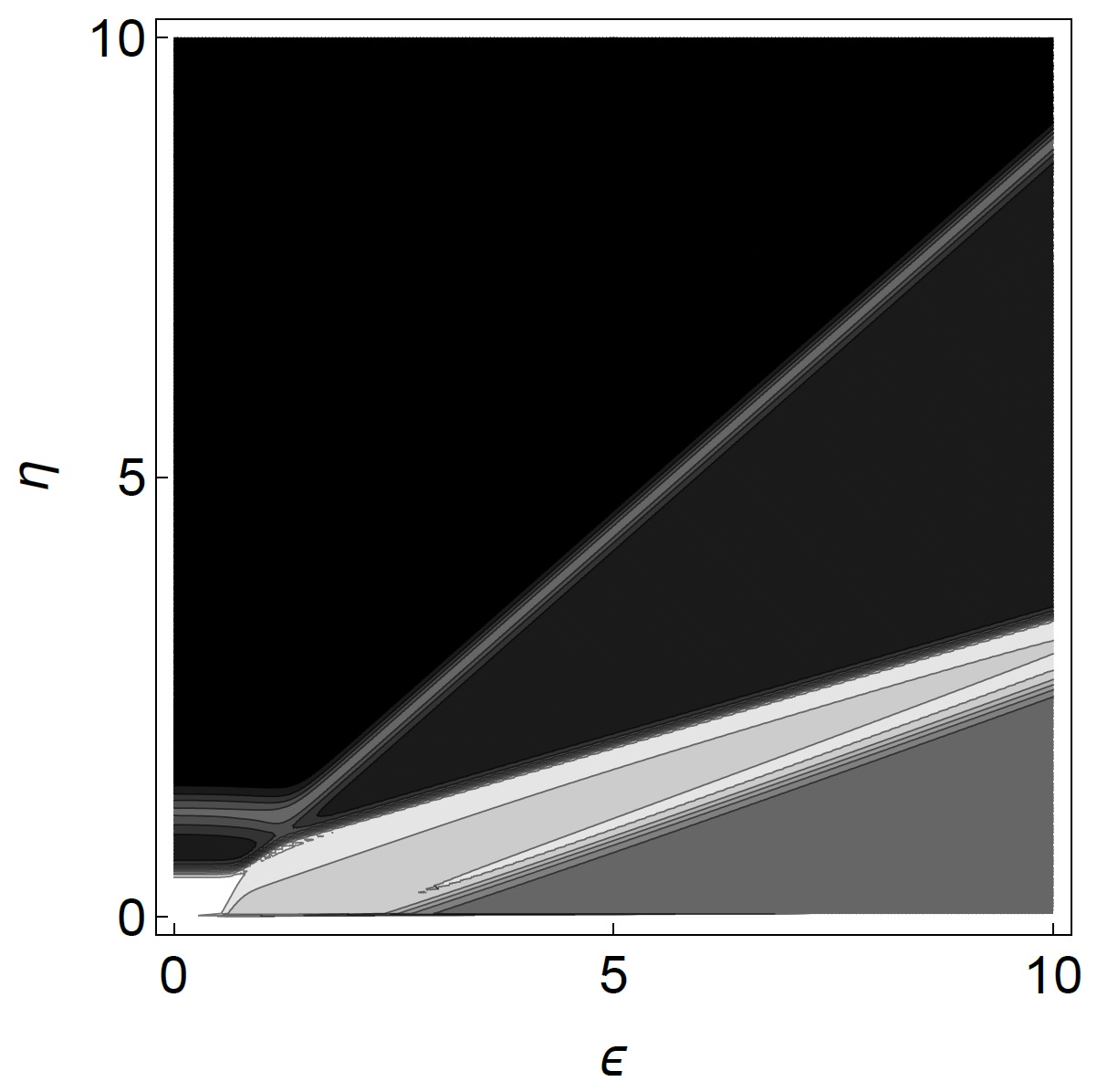}}\\
\subfigure[]{\includegraphics[width=0.40\textwidth, angle=0]{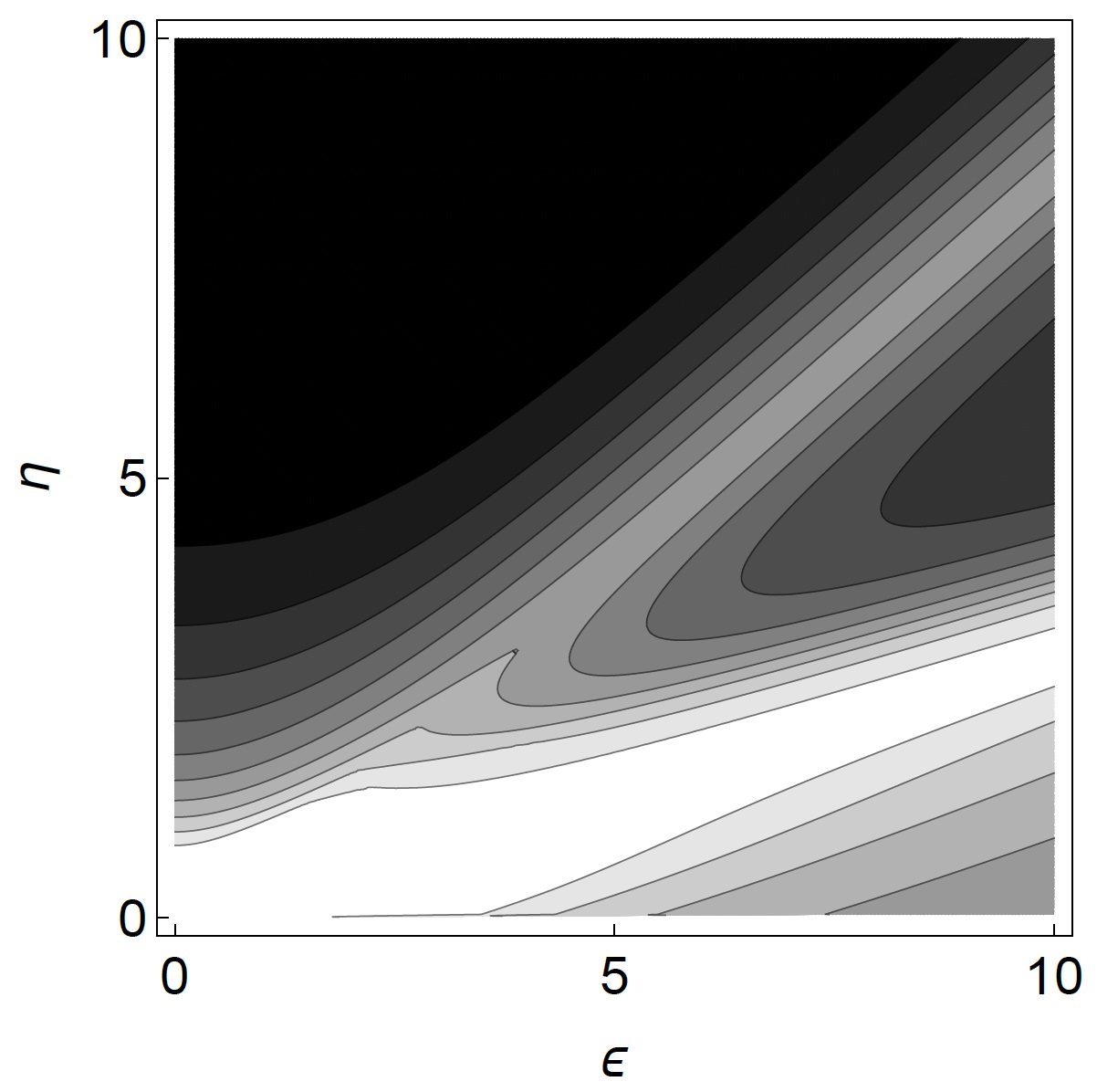}} \qquad
\subfigure[]{\includegraphics[width=0.40\textwidth, angle=0]{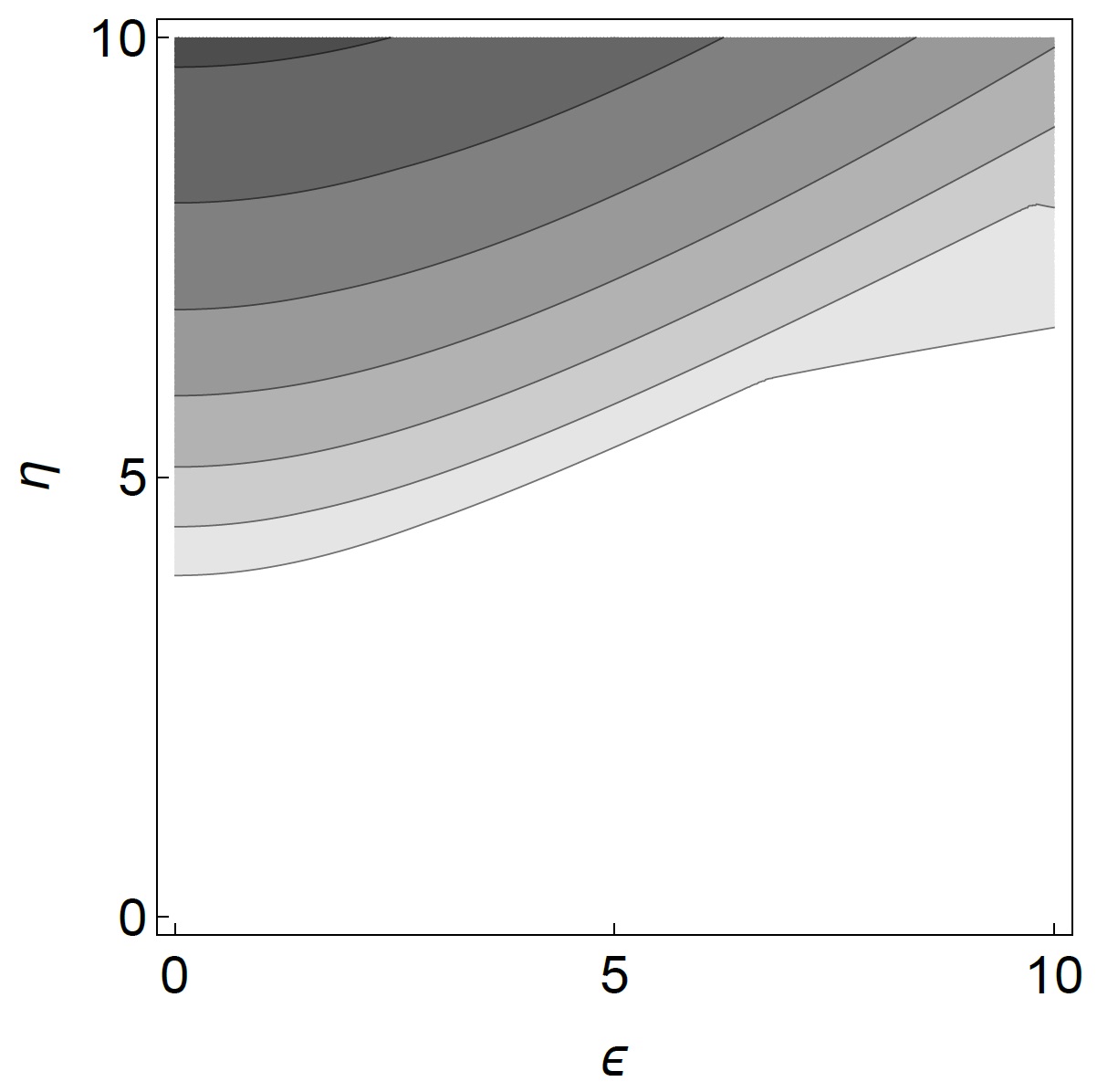}}
\includegraphics[width=0.08\textwidth, angle=0]{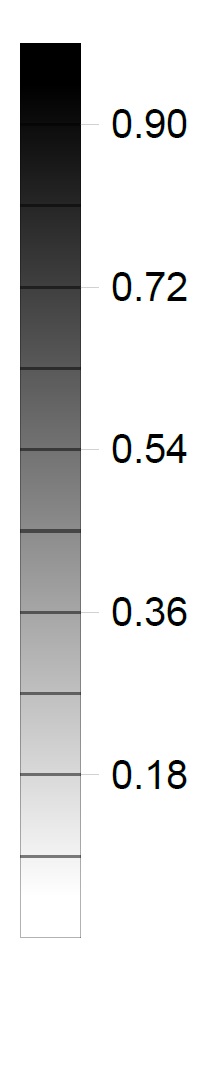}
\caption{Multipartite negativity of the reduced thermal state of the $N=4$ outer spins as a function of $\epsilon$ and $\eta$ (both in units of $\omega$), for $k_\mathrm{B} T / (\hbar\omega) = 0.01$ (a), $k_\mathrm{B} T / (\hbar\omega) = 0.1$ (b), $k_\mathrm{B} T / (\hbar\omega) = 1$ (c), $k_\mathrm{B} T / (\hbar\omega) = 5$ (d). (A darker grey corresponds to a higher value.)} \label{fig:TotalBehaviourN4}
\end{figure}

\begin{figure}
\subfigure[]{\includegraphics[width=0.40\textwidth, angle=0]{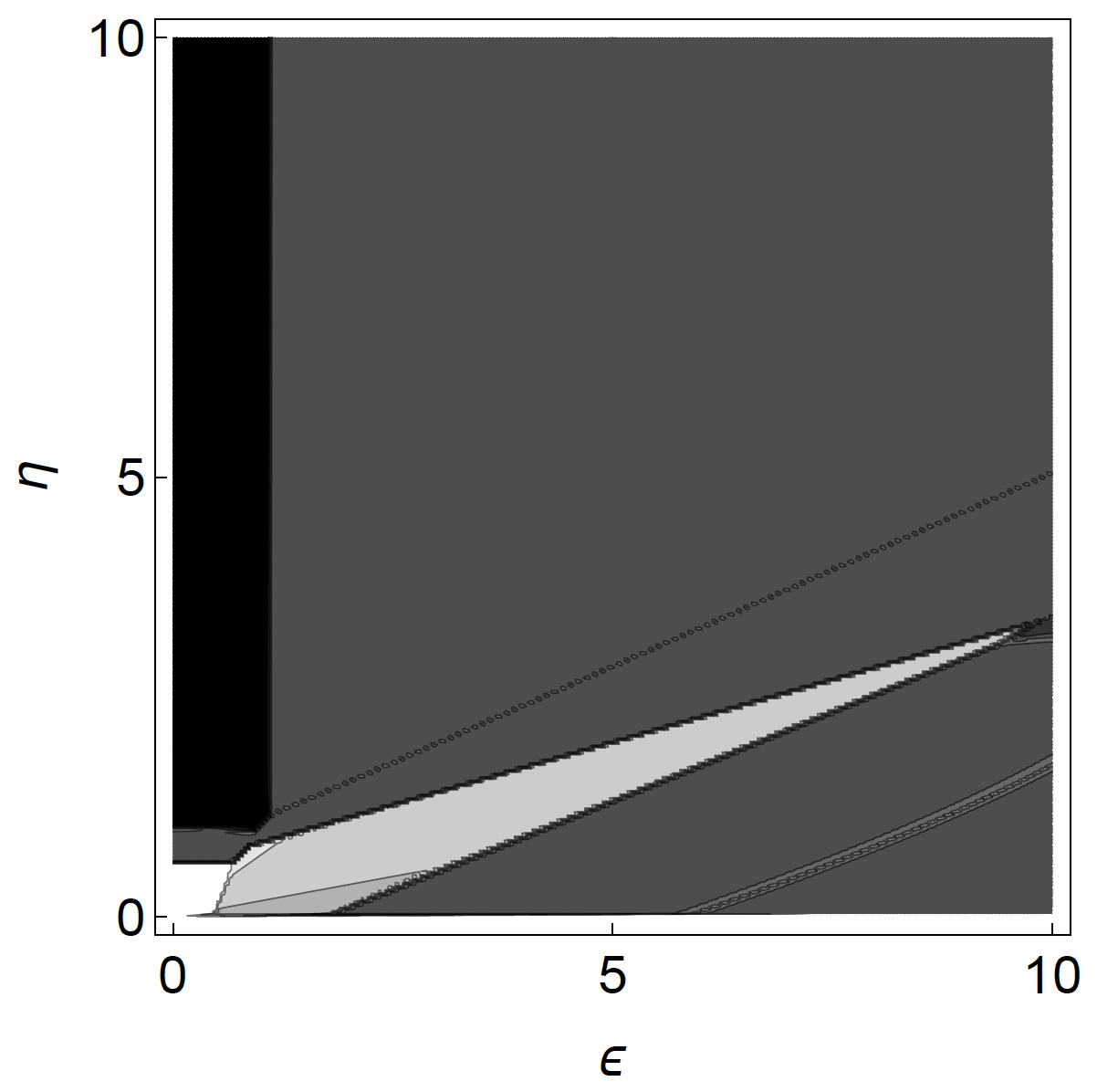}} \qquad
\subfigure[]{\includegraphics[width=0.40\textwidth, angle=0]{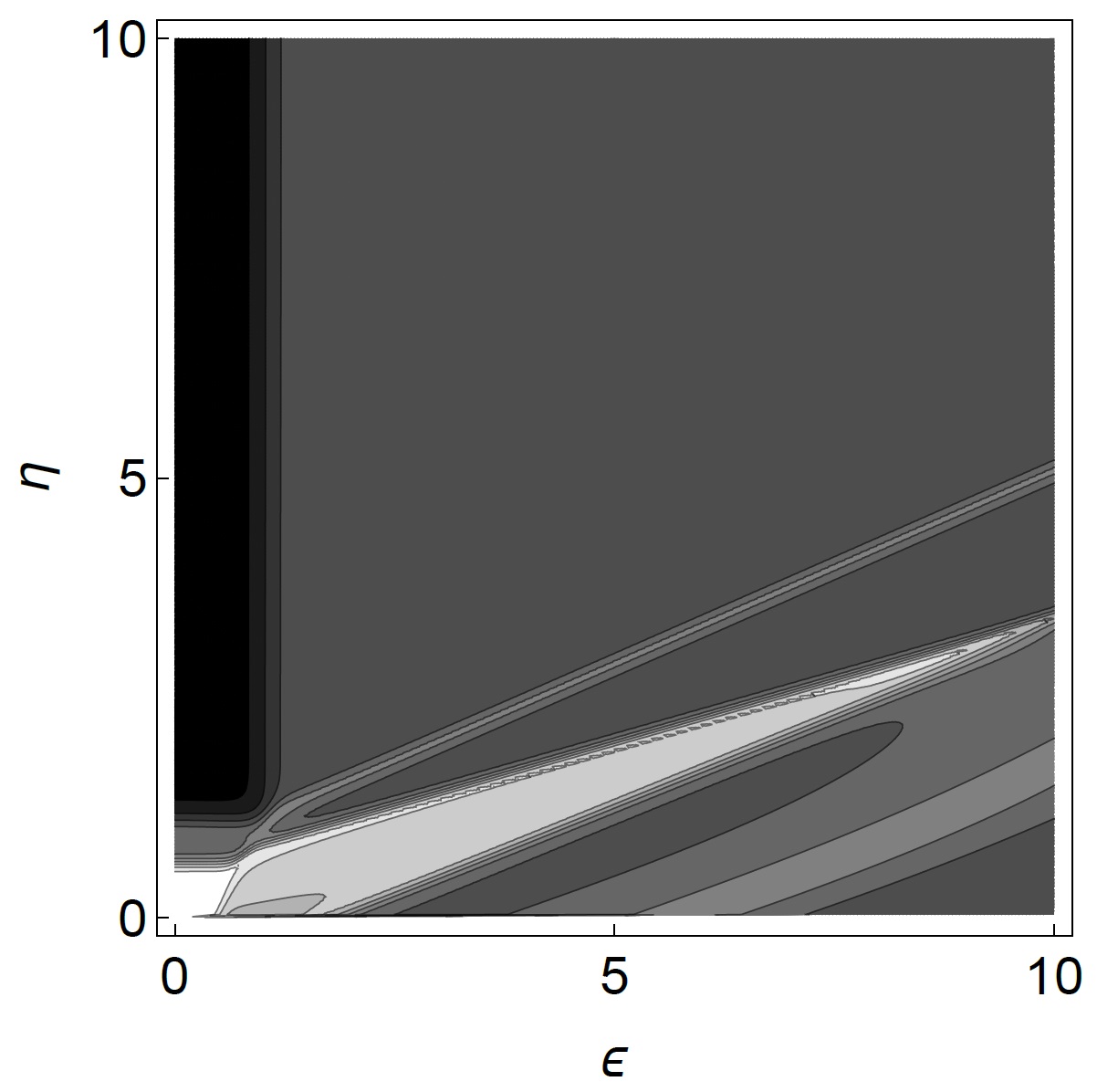}}\\
\subfigure[]{\includegraphics[width=0.40\textwidth, angle=0]{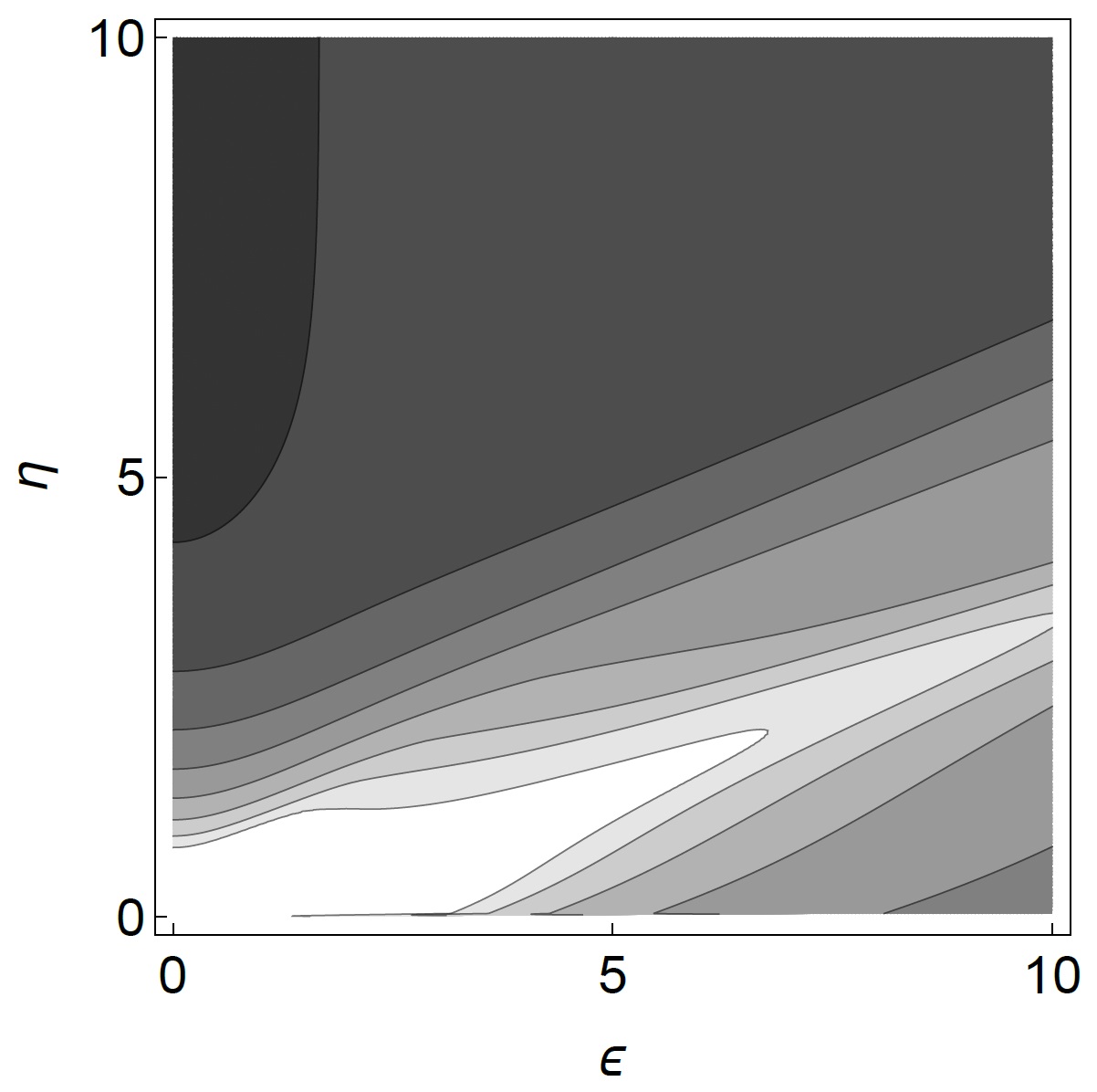}} \qquad
\subfigure[]{\includegraphics[width=0.40\textwidth, angle=0]{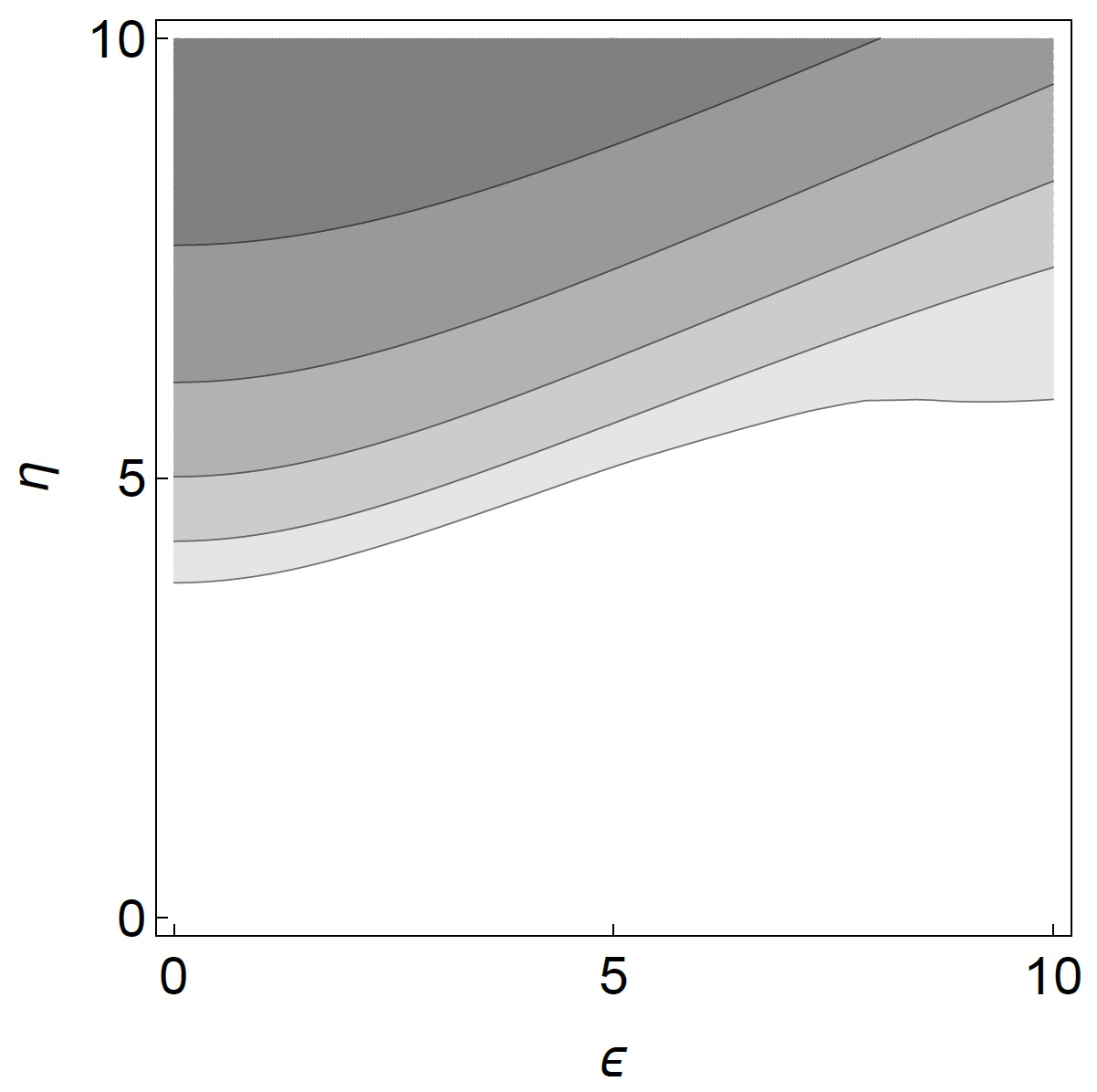}}
\includegraphics[width=0.08\textwidth, angle=0]{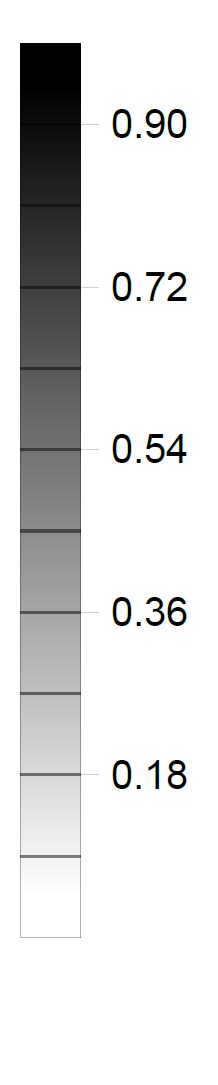}
\caption{Multipartite negativity of the reduced thermal state of the $N=5$ outer spins as a function of $\epsilon$ and $\eta$ (both in units of $\omega$), for $k_\mathrm{B} T / (\hbar\omega) = 0.01$ (a), $k_\mathrm{B} T / (\hbar\omega) = 0.1$ (b), $k_\mathrm{B} T / (\hbar\omega) = 1$ (c), $k_\mathrm{B} T / (\hbar\omega) = 5$ (d). (A darker grey corresponds to a higher value.)} \label{fig:TotalBehaviourN5}
\end{figure}

In Figs.~\ref{fig:TotalBehaviourN4} and \ref{fig:TotalBehaviourN5} are shown the numerical evaluations of the multipartite negativity in a wide range of values of the parameters $\eta$ and $\epsilon$, for different values of the temperature. Also in these cases, non monotonicity of the parameter $\mathcal{N}$ is well visible at relatively low temperatures.

\section{Conclusions}\label{sec:conclusion}

In this paper we have discussed the competition between two different couplings in the generation of thermal entanglement: one coupling is a direct interaction between three spins, while the second coupling is an interaction of each of the previous three spins with a fourth one. Since, generally speaking, both the interactions are responsible for the establishment of correlations between the three peripheral spins, one could expect a cooperation between them. Instead, our analysis clearly shows that in some cases it happens the opposite.

We have focused our analysis on the entanglement between the three peripheral spins when the whole system is in its thermal state, hence tracing over the degrees of freedom of the central spin. The two interactions, considered separately, are sources of thermal entanglement, but when they are both present they can compete, to the point that thermal quantum correlations can be completely cancelled, even at zero temperature. It does not happen for every values of the coupling constants, and this phenomenon disappears at (relatively) high temperature, but its occurrence is anyway an interesting fact which relies exactly on the interplay between the two interactions. This is well visible especially in the zero temperature diminishing that we have studied analytically, looking at the ground state of the system. Indeed, state $\Ket{\psi_1}$ is a linear combination of a completely factorized state and of a state involving a Werner state of the three peripheral spins, and the weights of such two contributions depend on how $\eta$, $\epsilon$ and $\omega$ compare to each other.

Finally, we have extended our analysis to the cases with $M>3$ outer spins, introducing the multipartite negativity as a natural generalization of the tripartite one. The numerical study of the multipartite negativity of the reduced thermal state of the outer spins for $M=4$ and $M=5$ shows in a clear way  the appearance of non monotonicity also in this new situation.



\end{document}